\documentstyle[prd,aps,preprint,epsfig]{revtex}

\tighten

\begin{document}
\draft
 
\pagestyle{empty}

\preprint{
\noindent
\hfill
\begin{minipage}[t]{3in}
\begin{flushright}
LBNL--46444 \\
UCB--PTH--00/24 \\
July 2000
\end{flushright}
\end{minipage}
}

\title{Selection rules in three-body B decay from factorization}

\author{
Mahiko Suzuki
}
\address{
Department of Physics and Lawrence Berkeley National Laboratory\\
University of California, Berkeley, California 94720
}


\maketitle

\begin{abstract}
Extending the dynamics underlying the factorization calculation of 
two-body decays, we propose simple selection rules for nonresonant
three-body $B$ decays.  We predict, for instance, that in the Dalitz plot 
of $B^0\rightarrow \overline{D}^0 \pi^+\pi^-$, practically no events 
should be found in the corner region of $E(\pi^+) \leq \Lambda_{QCD}$ 
as compared with the corner of $E(\pi^-) \leq \Lambda_{QCD}$. We also 
predict that there should be very few three-body decay events containing one 
soft meson resonance and two energetic mesons or meson resonances. 
The selection rules are quite different from the soft-pion theorem,
since they apply to different kinematical regions. For $B^0\rightarrow 
\overline{D}^0 \pi^+\pi^-$, the latter predicts that the decay matrix 
element vanishes in the zero four-momentum limit of $\pi^-$ instead 
of $\pi^+$. Since this marked difference from the soft-pion theorem 
is directly related to the issue of short-distance 
QCD dominance in two-body B decays, experimental test of the selection 
rules will shed light on strong interaction dynamics of $B$ decay. 
\end{abstract}

\pacs{PACS number(s) 13.20.He, 12.38.-t, 11.30.Rd, 11.30.-q}
\pagestyle{plain}
\narrowtext

\setcounter{footnote}{0}

\section{Introduction}
   The factorization calculation\cite{Stech} has successfully reproduced 
the decay rate for many two-body channels of $B$ decay. It has been 
argued\cite{BJ,Beneke} that the factorization should become exact in 
the large $b$-quark mass limit and that the only deviation from the 
factorization is short-distance corrections of $O(\Lambda_{QCD}/m_b)$.  
For some channels of $B$ decay, however, the factorization calculation 
appears to be in clear disagreement with current measurement\cite{Beneke2}.
Theoretically the factorization and the final-state interaction (FSI) 
are closely tied together\cite{Taiwan}. If the factorization is a good
approximation, the FSI phase should be small in general. If the FSI 
phase turns out to be large in experiment, it is a warning sign against 
the factorization. Some {\it initial state} interaction can also 
affect the factorization\cite{Taiwan,Gatto}. The observed smallness 
of the upper bounds on the FSI phases of the two-body $b\rightarrow c$ 
processes such as $B\rightarrow \overline{D}\pi$ is consistent 
with absence of long-distance FSI\cite{Bphase}. On the other hand,
purely phenomenological suggestions were recently made about possibility 
of large FSI phases in $B\rightarrow K\pi$, $\pi\pi$\cite{Hou}.  
At present, little is known in experiment about FSI phases of 
decays other than the charm producing channels.  

The factorization clearly fails in the $D$ decay in spite of the relatively 
small value of $\Lambda_{QCD}/m_c = 1/7\sim 1/5$. The observed relative 
FSI phase in $D\rightarrow \overline{K}\pi$ is close to 
$90^{\circ}$\cite{Dphase} though one would naively expect the 
factorization to work reasonably well if $\Lambda_{QCD}/m_c$ is 
really the controlling parameter of its corrections.\footnote{  
It has been well established that the FSI phase is close to $90^{\circ}$ 
in most of the two-body decay channels of $J/\psi$ so far
analyzed\cite{psiphase}.} In two-body $B$ decays the {\it c.m.} 
energy between a fast quark and a spectator grows only like 
$\sqrt{m_b\Lambda_{QCD}}$ with $m_b$, not linearly in $m_b$. Therefore,
$\sqrt{m_b\Lambda_{QCD}}$ must be large in order for perturbative QCD 
to be applicable. Even if one accepts the asymptotic validity of the 
factorization in theory\cite{Beneke}, one should test 
in experiment whether or not the $B$ mass of 5.3 MeV is high enough 
to be in the asymptopia of its final states.

In this paper, we extend the basic premise of the factorization and 
explore its consequences in three-meson decay channels of $B$ decay 
in which one of mesons is soft. We propose simple selection 
rules which will provide a further test of the fractorization 
at the $B$ mass scale. Our approach is purely phenomenological 
and we do not attempt here a rigorous justification of the selection
rules by perturbative QCD. When we compare the selection rules 
with the soft-pion theorem of chiral symmetry, it appears as if they 
were incompatible. The origin of this superficial incompatibility is traced
to difference in kinematics: The selection rules hold when the {\it c.m.}
energy $\sqrt{m_b\Lambda_{QCD}}$ between a fast quark and a soft quark is 
sufficiently large, while the soft-pion theorem holds when the same 
{\it c.m.} energy is equal to a light quark mass, {\it i.e.,} negligibly
small. The purpose of our selection rules is to probe if 
$\sqrt{m_b\Lambda_{QCD}}$ in $B$ decay is in the QCD asymptopia or not.

\section{Basis of factorization}
  We state the factorization of two-body decays as follows: 
When a $B$ meson decays with a current-current interaction through
the quark decay process,
\begin{equation}
  \overline{b} \rightarrow \overline{q}_1 + q_2 + \overline{q}_3, \label{qqq}
\end{equation}
either $\overline{q}_1q_2$ or $\overline{q}_3q_2$ forms one meson and the 
remaining $\overline{q_3}$ or $\overline{q}_1$ forms the other meson with 
a soft quark $q_4$ from the light spectator cloud around $\overline{b}$. 
(See Fig.1a.) The formation of a meson by two hard quarks through a 
color-singlet current is given by a decay constant while the formation of 
the other meson is described by a transition form factor of the other 
current. It is important that once combinations of $q\overline{q}$ 
have been chosen for meson formation, no flavor exchange occurs 
thereafter, not even a charge exchange. Any flavor-changing process is 
supposedly a higher order process of perturbative QCD. Therefore the final 
mesons retain their flavor contents, including charges, determined by 
the short-distance quark decay process. This leads us to various 
constraints of the factorization which relate different isospin 
amplitudes of two-body decays for a given decay operator.

What happens if three mesons are produced ? In the case that all final 
mesons are energetic, each meson should contain one of the three final 
quarks of Eq.(\ref{qqq}) picking up one soft quark from the spectator cloud. 
(See Fig.1b.) In the case that two of the mesons are energetic and one 
is very soft, the energetic mesons are formed just as in the two-body 
decay and the third one emerges from the spectator cloud with 
energy no larger than $O(\Lambda_{QCD})$. In this case the energetic 
mesons retain the history of the weak interaction while the soft meson 
is made of whatever left in the spectator cloud. This is the natural 
extension of the factorization from two-body decays to three-body 
decays.  We call this picture as the factorization for three-body 
decays and explore its consequences here. The {\it c.m.} energy of
the FSI between an energetic meson and the soft cloud is  
$O(\sqrt{m_b\Lambda_{QCD}})$, which is the same as the {\it c.m.} energy of
$\overline{q}_1$ and $q_4$ of the two-body decay in Fig.1a.\footnote{
The {\it c.m.} energy between the soft meson and the soft quark in the 
fast meson is $O(\Lambda_{QCD})$, but our selection rules do not depend
on a soft interaction between them.} If $\sqrt{m_b\Lambda_{QCD}}$ 
is large and short-distance QCD dominates in FSI, flavor exchange 
between any pair of final mesons would be a higher order correction. 
In this case, the three quarks produced by the $\overline{b}$-quark 
can be tracked down into the flavor contents of the final mesons. 

\section{Illustration by $B^0\rightarrow \overline{D}^0\pi^+\pi^-$}
We illustrate derivation of our selection rule in the decay,
\begin{equation}
    B^0(\overline{b}{\bf d})\rightarrow \overline{c}u\overline{d}{\bf d}
                      \rightarrow \overline{D}^0(\overline{c}u)
                                  \pi^+(\overline{d}u)
                                  \pi^-(\overline{u}d),  \label{Dpipi}
\end{equation}
which proceeds through the tree-operator
$(\overline{b}c)(\overline{u}d)$ and its gluon corrections. The boldface 
letter ${\bf d}$ in Eq.(\ref{Dpipi}) represents a cloud of light 
quark-antiquarks and gluons around $\overline{b}$ which carries the net 
flavor of the $d$-quark;
\begin{equation}
  {\bf d} =  d + d(\overline{u}u) + d(\overline{d}d) + \cdots. \label{spec}
\end{equation}
We refer to ${\bf d}$ as the spectator $d$-quark cloud. When the decay 
products $\overline{c}u\overline{d}$ of $\overline{b}$ enter three 
mesons one each, all mesons are energetic. In this case we cannot 
find a simple flavor selection rule of the factorization. The reason 
is as follows: Even if no flavor exchange occurs between final mesons, 
the process $\overline{u}u\leftrightarrow\overline{d}d$ occurs perpetually 
inside the cloud prior to meson formation. This annihilation process
has the same effect as a flavor exchange between final mesons since each 
meson picks up one soft quark from the cloud. Therefore, no simple 
testable selection rule results in this case. 

On the other hand, if one of the pions is soft, it must be generated by
the spectator cloud without involving a hard quark. The three  
quarks from $\overline{b}\rightarrow\overline{c}u\overline{d}$ must 
go into $\overline{D}^0$ and the energetic $\pi$.\footnote{
A quantitative estimate will be given later to show that 
the soft tails of $u$ and $\overline{d}$ are negligible.}
The hard $u$ can form either $\pi^+$, $\pi^0$, or $\overline{D}^0$ 
while the hard $\overline{d}$ can form either $\pi^+$ or $\pi^0$.
Even though $\overline{d}$ or $\overline{c}$ picks one soft quark from
the cloud, the resulting meson is energetic. After one quark ($u$ or 
$d$) is removed from the spectator cloud {\bf d} of total charge $-1/3$, 
the remainder of the cloud is either negatively charged or neutral. That is, 
the soft pion can be either $\pi^-$ or $\pi^0$, but not $\pi^+$. We thus
conclude that $\pi^+$ cannot be soft in $B^0\rightarrow
\overline{D}^0\pi^+\pi^-$. This selection rule is depicted in Fig.2. 

In fact, the selection rule holds irrespective of the decay interaction.
It depends only on the flavor content of the spectator {\bf d}: 
{\it When one quark is taken away from a cloud of charge $-1/3$, 
the remainder cannot be positively charged.} The key is that mesons 
carry quantum numbers of quark-antiquark no matter what their 
Fock-space compositions are. The only assumption we have made here 
is that the energetic mesons do not exchange a flavor
with the soft meson by short-distance QCD dominance in FSI.

Our prediction should be tested in the form,
\begin{equation}
  B(\pi^+_{soft}\overline{D}^0\pi^-)/B(\pi^-_{soft}\overline{D}^0\pi^+) = 0,
                                                                 \label{sel}
\end{equation}
where $|{\bf p}_{\pi}|\leq O(\Lambda_{QCD})$ for the soft pion. The bands of 
the $\pi\pi$ and $\overline{D}^0\pi$ resonances should be excluded in testing 
Eq.(\ref{sel}). If, contrary to the factorization, elastic or inelastic 
rescattering is important at $E_{\pi\pi} = \sqrt{m_b\Lambda_{QCD}}$ 
($\simeq 1.5$ GeV), $\pi^+_{soft}\overline{D}^0\pi^-$ can be 
fed through the unsuppressed decay modes, for instance, through
$\pi^0_{soft}\overline{D}^0\pi^0$ followed by the forward charge exchange 
rescattering $\pi^0\pi^0\rightarrow\pi^+\pi^-$ or through $\pi^-_{soft}
\overline{D}^0\pi^+$ followed by the backward elastic rescattering
$\pi^+\pi^-\rightarrow\pi^-\pi^+$, {\it i.e.,} the forward rescattering 
of two units of charge exchange. Thus the selection rule is sensitive 
to FSI.\footnote{To be precise, rescattering to individual channels
need not be small if they are cancelled out by destructive interference.}
The FSI of our concern is the soft-hard process at {\it c.m.} energy of 
$O(\sqrt{m_b\Lambda_{QCD}})$ that involves a hard quark, since the soft-soft 
FSI is no more than a different pick of soft quarks from the cloud. 

Selection rules similar to Eq.(\ref{sel}) can be derived for many other 
three-body decay modes. Considering their relative simplicity, it is 
worthwhile exploring such selection rules for better
understanding of strong interaction dynamics in $B$ decay. 

\section{Generalization}
\subsection{Soft pion and kaon production}
The $\overline{D}^0$ meson can be replaced by any meson or meson resonance in
Eq.(\ref{Dpipi}); $B^0\rightarrow \pi^+ \pi^- M^0$, where $M^0$ denotes 
a neutral meson or meson resonance such as $K^0$, $\pi^0$, or $\rho^0$. 
The selection rule reads:
\begin{equation}
 B(B^0\rightarrow\pi^+_{soft}\pi^-M^0)/B(B^0\rightarrow\pi^-_{soft}\pi^+M^0) 
     = 0. 
\end{equation}
We can apply the same reasoning to $B^+$ by replacing 
the spectator ${\bf d}$ by ${\bf u}$;
\begin{equation}
 B(B^+\rightarrow\pi^-_{soft}M^+\pi^+)/B(B^+\rightarrow\pi^+_{soft}M^+\pi^-)
  = 0,
\end{equation} 
where $M^+$ is a positively charged meson or meson resonance such as
$K^+$, $\pi^+$, and $\rho^+$.
For $B_s$, we can write the selection rule
\begin{equation}
             B(B_s\rightarrow K_{soft}\overline{K}M)/B(
               B_s\rightarrow\overline{K}_{soft}KM)
                 = 0.
\end{equation}

If we include $\overline{s}s$ in the spectator cloud, we obtain 
corresponding selection rules,
\begin{eqnarray}
  B(B^+\rightarrow K^0_{soft}\overline{K}M)/B(
        B^+\rightarrow K^+_{soft}\overline{K}M)  &=& 0, \\ \label{kaon1}
  B(B^0\rightarrow K^+_{soft}\overline{K}M)/B(
        B^+\rightarrow K^0_{soft}\overline{K}M)  &=& 0. \label{kaon2}
\end{eqnarray}                                    
Furthermore,
\begin{eqnarray}
 B(B^+\rightarrow\overline{K}_{soft}KM)/B(
   B^+\rightarrow K^+_{soft}\overline{K}M) &=& 0,\\ \label{kaon3}
 B(B^0\rightarrow\overline{K}_{soft}KM)/B(
   B^0\rightarrow K^0_{soft}\overline{K}M) &=& 0. \label{kaon4}
\end{eqnarray}
Since the rest energy of a kaon exceeds 
$\Lambda_{QCD}$, formation of any kaon from the spectator cloud is 
kinematically suppressed. Eqs.(8) to (\ref{kaon4}) mean that the 
factorization-forbidden modes are even further suppressed than the
modes in the denominators.

\subsection{Soft resonance production in quasi-three-body decay}
The lightest meson resonance that the spectator cloud can produce
is the $\rho$-meson. Since its rest mass far exceeds $\Lambda_{QCD}$, 
$B\rightarrow\rho_{soft}M_1M_2$ is suppressed no matter what charge 
state the $\rho$-meson is in. The cloud simply does not have 
enough energy to produce $\rho$ even at rest.   

To be concrete, consider $B\rightarrow\rho\pi M$. Then the lack of 
energy in producing a resonance from the cloud leads us to
\begin{eqnarray}
  B(B^+\rightarrow\rho_{soft}M\pi^+)/B(B^+\rightarrow\pi^+_{soft}M\rho) 
                                           &=& 0, \nonumber \\
  B(B^0\rightarrow\rho_{soft}M\pi^-)/B(B^0\rightarrow\pi^-_{soft}M\rho)
                                           &=& 0,   \label{selrho}
\end{eqnarray}
where $\rho$ can be in any charge state. In Eq.(\ref{selrho}) $\rho$ 
may be replaced by $\omega$ or some other resonance. Similarly,
\begin{equation}
  B(B_s\rightarrow \overline{K}^*_{soft}K M)/B(
    B_s\rightarrow \overline{K}_{soft}K^* M) = 0.
\end{equation}

Though we have listed only the selection rules that are relatively simple
and clean, one can write down many more selection rules by the same
reasoning. 

Since meson or meson resonance production of energy higher than 
$O(\Lambda_{QCD})$ from the cloud is highly suppressed kinematically, 
we may raise the softness limit considerably from $|{\bf p}|\leq
\Lambda_{QCD}$ to, for instance, $|{\bf p}|\leq m_{\rho}$ in actual 
testing of the selection rules, provided that one can avoid resonance 
bands of quasi-two-body decays. If we raise it too high, however, the 
tail of the energy spectrum of a decay product quark enters the ``soft 
meson''. To find a compromise, we need some numerical study. 
 
\section{How soft should the soft pion be ?}

Our selection rules are based on the postulate that a soft meson 
must be produced entirely from the spectator cloud. Therefore, 
they would fail if a soft meson contains a significant contribution 
from the low-energy tail of a decay product of $\overline{b}\rightarrow
\overline{q}_1q_2\overline{q}_3$. For this reason we would like to know 
how much the low-energy tail contributes when we define the soft meson by
$|{\bf p}_{\pi}|\leq\Lambda_{QCD}$. If the tail is indeed negligible, then 
we should ask how much we can raise the upper limit of softness in actual 
test without violating the selection rules substantially.
 
To be concrete, consider $B^0\rightarrow\pi^+_{soft}\overline{D}^0\pi^-$ 
again. Suppose that we define the soft pion by its momentum
$|{\bf p}_{\pi}|\leq p_{max}$.
Estimate of the matrix element involves uncertainties in soft meson
formation. If we use the quark-hadron duality for the low-energy tail 
region and replace the meson wavefunctions by their average 
values, we can obtain a crude estimate of the decay rate ratio as
\begin{eqnarray}
  \frac{\Gamma(\pi^+_{soft}\overline{D}^0\pi^-)}{\Gamma(
               \pi^-_{soft}\overline{D}^0\pi^+)} &\approx &
      C^2 \frac{\int_{|{\bf p}_u|\leq p_{max}}
     d\Gamma(\overline{b}\rightarrow \overline{c}u\overline{d})}{
             \int_{|{\bf p}_u|\geq m_b/2-p_{max}}^{m_b/2}
     d\Gamma(\overline{b}\rightarrow \overline{c}u\overline{d})} \\ \label{q0}
     & \simeq & \epsilon^2C^2,  \label{q1}
\end{eqnarray}
where $\epsilon = 2p_{max}/m_b$. 
The factor $C$, which arises from difference in meson formation probability, 
is given by the ratio of the pion wavefunction $\Phi(x)$ on the 
light-cone as
\begin{equation}
    C\simeq \frac{\Phi(\epsilon)\Phi(x_0)}{\Phi(1/2)^2},
\end{equation} 
where $x_0=\Lambda_{QCD}/(p_{max}+\Lambda_{QCD})$. The value of $C$ is 
sensitive to $\Phi(\epsilon)$ near the end point. Let us extrapolate 
$\Phi(x)\propto x(1-x)$ smoothly to $x=0$. Then for $p_{max}=\Lambda_{QCD}$
\begin{equation}
 \frac{\Gamma(\pi^+_{soft}\overline{D}^0\pi^-)}{\Gamma(
              \pi^-_{soft}\overline{D}^0\pi^+)}
         \approx 16\biggl(\frac{2\Lambda_{QCD}}{m_b}\biggr)^4. \label{q2} 
\end{equation}
The right-hand side is a fraction of 1\% for $\Lambda_{QCD} = 300$ MeV. 
It is indeed small due to the phase space suppression of the soft 
$u$-quark and a smaller meson formation probability for 
a quark pair of highly uneven energies.  From this estimate we expect 
that accuracy of our selection rules should be very high.
When $p_{max}$ is raised to $m_{\rho}$, the ratio in Eq.(\ref{q1}) 
goes up to 8\%. In carrying out experimental test, therefore, we may 
be able to raise the softness upper limit from $\Lambda_{QCD}$ to 
$m_{\rho}$. The main background in testing our this selection rule
will be a spillover from the resonance bands of $\rho^0$ and $D^{*-}$ 
due to errors in pion measurement.

\section{Comparison with the soft-pion theorem}

In the zero four-momentum limit of the soft pion, we can relate
three-body decay matrix elements to two-body decay matrix elements 
by the soft-pion theorem of chiral symmetry. It can
be expressed with $f_{\pi}=130$ MeV as
\begin{equation}
    M(B\rightarrow\pi^-(k_{\mu})\overline M_1M_2)
    \stackrel{k_{\mu}\rightarrow 0}{=} \frac{1}{f_{\pi}}
   \langle M_1M_2|[u^{\dagger}\gamma_5d, H_w(0)]|B\rangle +k_{\mu}M^{\mu},
\end{equation}
where the $k_{\mu}M^{\mu}$ term gives the correction to
the soft-pion limit\cite{AD}. For soft $\pi^+$ emission, 
$u^{\dagger}\gamma_5d$ should be replaced by $d^{\dagger}\gamma_5u$. 
Upon extrapolation to the physical pion at rest $k_{\mu} = 
(m_{\pi},{\bf 0})$, the correction term $k_{\mu}M^{\mu}$ is of 
$O(k_{\mu}/\Delta m)$ where $\Delta m$ is a typical excitation energy 
of the $B$ system. According to the traditional wisdom, the Born 
diagram of $B^*(1^-)$ is likely the most important correction 
(see Fig.3a) since it is nearly degenerate with $B$ 
($\Delta m\simeq 46$ MeV). This correction can be of $O(1)$ or even 
dominate over the zero-four-momentum limit for a physical soft pion. 
It is easy to see that the $B^*$ contribution to $k_{\mu}M^{\mu}$ 
satisfies the selection rule Eq.(\ref{sel}).   
In fact, this is the case for the contributions from all Born diagrams. 
The reason is that the quark diagrams for the Born diagrams of Fig.3a are 
topologically equivalent to those of the three-body factorization process.
The only difference between them is whether a soft pion is emitted from 
the cloud before or after weak interaction takes place. 

If we keep only the soft-pion limit, we obtain
\begin{eqnarray}
  \lim_{k_{\mu}\rightarrow 0}M(\pi^+(k_{\mu})\overline{D}^0\pi^-) 
         &=&\frac{1}{f_{\pi}}M(\overline{D}^0\pi^0), \label{s1} \\ 
  \lim_{k_{\mu}\rightarrow 0}M(\pi^-(k_{\mu})\overline{D}^0\pi^+) &=& 0.
                                                     \label{s2} 
\end{eqnarray}
Eq.(\ref{s2}) results from the vanishing commutator,
$[u^{\dagger}\gamma_5d, H_w]=0$. 
In contrast, the zero four-momentum limit of the 
factorization-forbidden decay $B^0\rightarrow\pi^+(k_{\mu})
\overline{D}^0\pi^-\;(k_{\mu}\rightarrow 0)$ is related to 
$B^0\rightarrow\overline{D}^0\pi^0$. While $M(\overline{D}^0\pi^0)$ is 
$O(1/N_c)$ for ${\cal O}_2$, it is the leading term for ${\cal O}_1$. 

Eqs.(\ref{s1}) and (\ref{s2}) look puzzling at the first sight. 
If we work out the soft-pion 
limit for $B^+\rightarrow\pi^+\pi^-\pi^+$, our three-body factorization 
predicts 
\begin{equation}
  \frac{B(B^+\rightarrow\pi^-_{soft}\pi^+\pi^+)}{B
         (B^+\rightarrow\pi^+_{soft}\pi^-\pi^+)}= 0,
\end{equation}
while the soft-pion theorem gives for the tree-operator ${\cal O}_1
=(\overline{b}_L\gamma_{\mu}d_L)(\overline{u}_L\gamma^{\mu}u_L)$ and 
${\cal O}_2=(\overline{b}_L\gamma_{\mu}u_L)(\overline{u}_L\gamma^{\mu}d_L)$
\begin{eqnarray}
  \lim_{k_{\mu}\rightarrow 0}M(B^+\rightarrow\pi^-(k_{\mu})\pi^+\pi^+) &=&
    \frac{2\sqrt{2}}{f_{\pi}}M(B^+\rightarrow\pi^+\pi^0),\\ 
  \lim_{k_{\mu}\rightarrow 0}M(B^+\rightarrow\pi^+(k_{\mu})\pi^-\pi^+) &=&
    \frac{1}{f_{\pi}}M(B^0\rightarrow\pi^0\pi^0).
\end{eqnarray}
The situation looks equally bizzare in $B\rightarrow K\pi\pi$ 
through the penguin operators. Since the penguin operators of 
$\overline{b}\rightarrow\overline{s}$ are all singlets of chiral
SU(2)$\times$SU(2), their commutators with the axial isospin charges
are zero. That is, the soft-pion limit should disappear 
for all matrix elements of $B\rightarrow\pi(k_{\mu})K\pi$ as $k_{\mu}
\rightarrow 0$ for the penguin operators.

It appears as if the selection rules contradicted the soft-pion theorem of 
chiral symmetry. We can trace the origin of this apparent discrepancy to the 
{\it c.m.} energy between a fast quark ($p_{\mu}$) from $\overline{b}$-decay 
and a soft quark ($k_{\mu}$) in the cloud. In the {\it physical} soft pion 
limit where our selection rules should hold,
\begin{equation}
     (p+k)^2 \simeq 2(k\cdot p) = O(m_b\Lambda_{QCD}),  \label{CM1}
\end{equation}   
while in the zero four-momentum limit
\begin{equation}
     (p+k)^2 \rightarrow p^2 = m_q^2. \label{CM2}
\end{equation}
In order for the factorization to work, the right-hand side of Eq.(\ref{CM1})
must be large enough to be in the asymptopia of QCD. The same variable 
is equal to the light quark mass in the soft-pion theorem. Their predictions
apply to different kinematical regions. 

The chiral Lagrangian has been successful for which the soft-pion 
theorem gives the leading terms. However, extrapolation to the physical 
region of $B$-decays is a long way because of large $B$ mass (cf
Eqs.(\ref{CM1}) and (\ref{CM2})) and involves large incalculable corrections.
 
We can understand what physical process the soft-pion theorem 
represents\cite{AD}. The commutator of the axial charge $Q_5^a = 
q^{\dagger}\gamma_5(\tau_a/2)q\; (a=1,2,3)$ with weak 
decay operator ${\cal O}_i$ arises from its commutator with the 
currents in ${\cal O}_i$. In the case of $B^0\rightarrow
\pi^+(k_{\mu})\overline{D}^0\pi^-$ at $k_{\mu}\rightarrow 0$ with 
${\cal O}_1 = (\overline{b}_L\gamma_{\mu}d_L)(\overline{u}_L\gamma^{\mu}c_L) 
=(\overline{b}_{L\alpha}\gamma_{\mu}c_{L\beta})(
\overline{u}_{L\beta}\gamma^{\mu}d_{L\alpha})$, 
for instance, the commutator with 
$\overline{u}_{L\beta}\gamma^{\mu}d_{L\alpha}$ 
survives to give 
\begin{equation}
   [d^{\dagger}\gamma_5u,{\cal O}_1] =
  (\overline{b}_{L\alpha}\gamma_{\mu}c_{L\beta})(
   \overline{u}_{L\beta}\gamma^{\mu}u_{L\alpha}
  -\overline{d}_{L\beta}\gamma^{\mu}d_{L\alpha}),   
\end{equation}
where the right-hand side is the $I_3=0$ partner of ${\cal O}_1$.
This commutator describes production of states of charge $Q=+1$ through 
$\overline{u}_{L\beta}\gamma^{\mu}d_{L\alpha}$ of ${\cal O}_1$ 
and subsequent fragmentation of the soft $\pi^+$ from them. 
Emission of a zero-four-momentum pion from the energetic 
quark pair $\overline{u}_{L\beta}\gamma^{\mu}d_{L\alpha}$  
is not suppressed by asymptotic freedom since 
no large momentum transfer occurs. After the $\pi^+$ emission, these states 
($\sim \overline{u}_{\beta}u_{\alpha}-\overline{d}_{\beta}d_{\alpha}$) 
and the operator $\overline{b}_{L\alpha}\gamma_{\mu}c_{L\beta}$ 
together produce the final state $\overline{D}^0\pi^-$. (See Fig.3b.)  
For soft $\pi^-$ emission, the commutator vanishes so that there is no 
diagram corresponding to Fig.3b. Thus physics in the zero-four-momentum
limit is quite different from what we have introduced 
as an extension of the factorization in this paper.

An important question is which picture gives a better description of 
the three-body $B$ decay in which one pion is soft. Is the approximation 
of $\sqrt{m_b\Lambda_{QCD}}\rightarrow\infty$ better  
than the zero four-momentum approximation ?  The issue should be 
settled by experimental test of the selection rules proposed here. 
  
\section{Summary}
   The dynamics underlying the factorization can be tested 
with the selection rules in three-body decays in which one of 
mesons is soft. The selection rules test whether the soft-hard FSI at 
{\it c.m.} energy of $O(\sqrt{m_b\Lambda_{QCD}})$ is in the QCD asymptopia or 
not. The rules are intuitive and do not depend on decay operators 
nor on form factors. If the selection rules are verified in experiment, we 
would obtain better understanding of physics underlying the factorization
and strong interactions in $B$ decay in general.

\acknowledgements
This work was supported in part by the Director, Office of Science, 
Division of High Energy and Nuclear Physics, of the
U.S.  Department of Energy under Contract DE--AC03--76SF00098 and in
part by the National Science Foundation under Grant PHY--95--14797.


\noindent
\begin{figure}
\epsfig{file=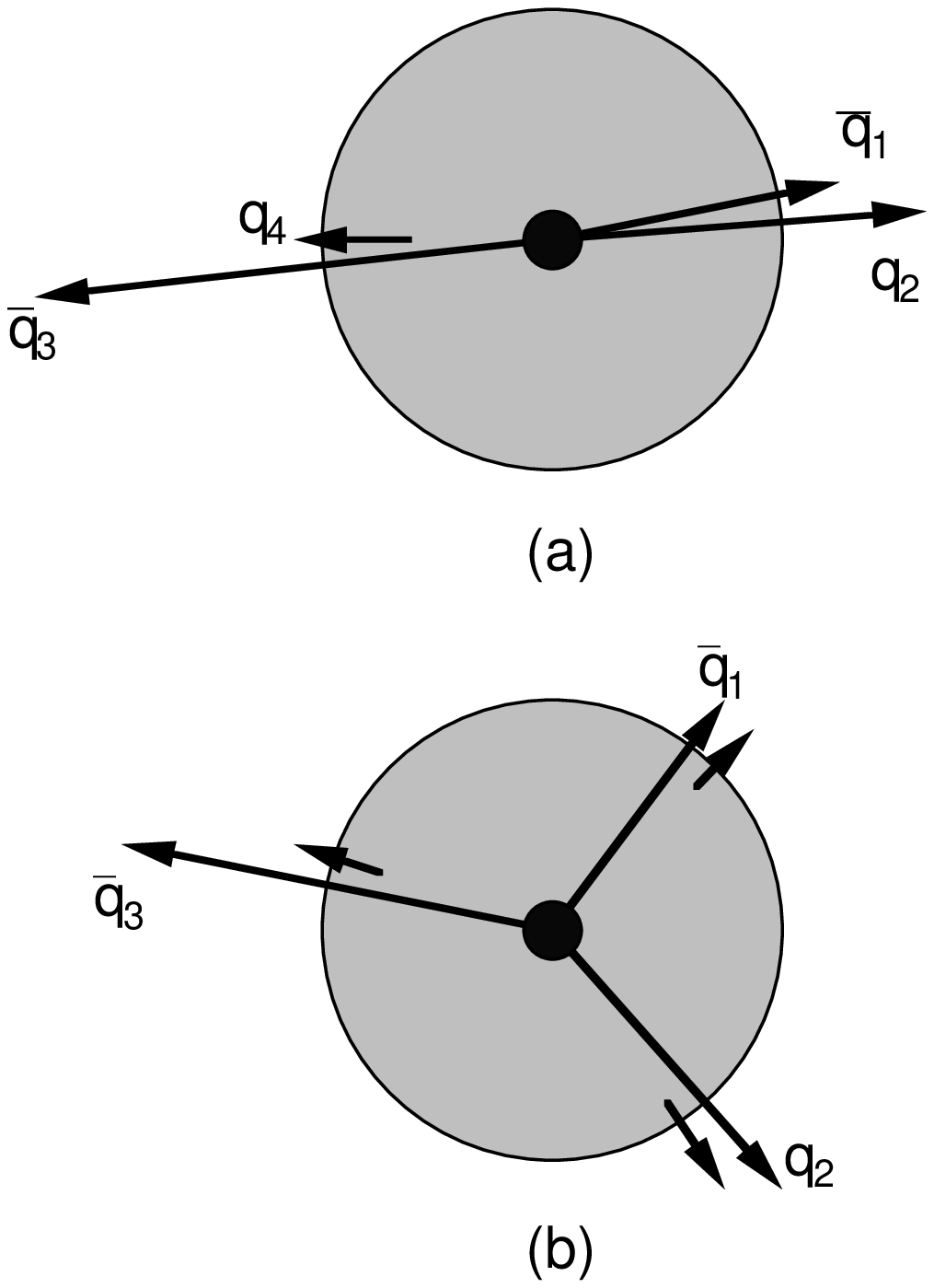,width=6cm,height=8cm}
\caption{(a) The quark picture of the factorization for a two-meson decay. 
(b) The extension to a three-meson decay in which all three mesons are
energetic. The shaded circle represents the spectator cloud, and 
the short arrows denote soft quarks from the cloud. When one of mesons
is very soft in a three-body decay, the decay occurs like (a) with 
the residual cloud turning into the soft meson. \label{fig:1}} 
\end{figure}

\noindent
\begin{figure}
\epsfig{file=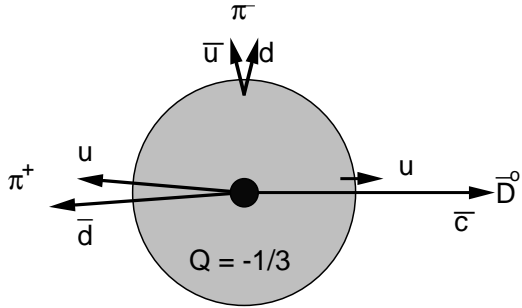,width=7cm,height=4.3cm}
\caption{The decay $B^0\rightarrow\pi^-_{soft}\overline{D}^0\pi^+$.
The decay $B^0\rightarrow\pi^0_{soft}D^-\pi^+$ is also allowed. \label{fig:2}}
\end{figure}

\noindent
\begin{figure}
\epsfig{file=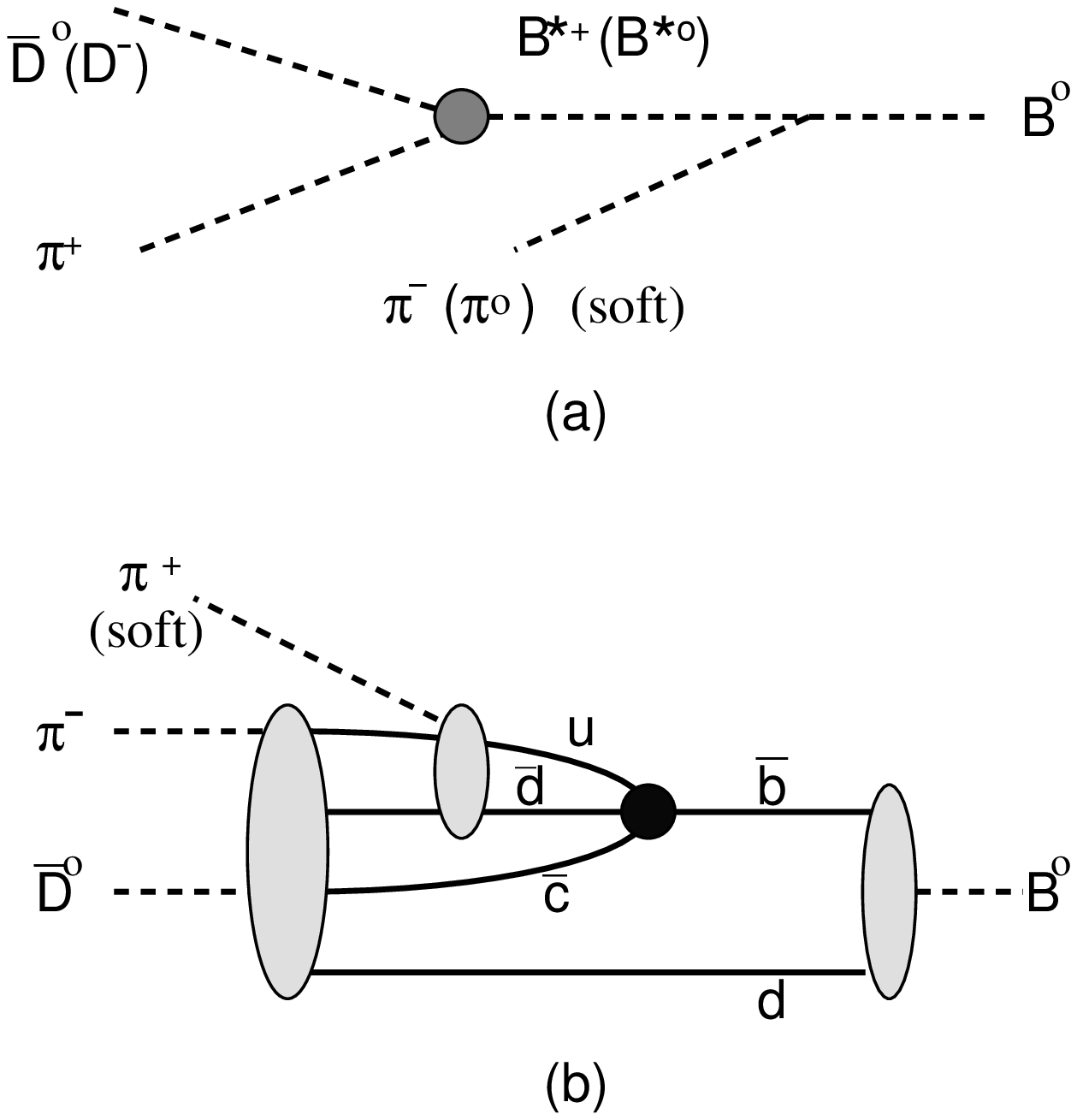,width=7cm,height=6.5cm}
\caption{(a) The leading contribution to 
$k_{\mu}M^{\mu}$ of $B^0\rightarrow\overline{D}^0\pi^+\pi^- (D^-\pi^+\pi^0)$  
when $\pi^- (\pi^0)$ is soft. It does not contribute to soft $\pi^+$ emission.
(b) The zero four-momentum limit of $\pi^+$ in 
$B^0\rightarrow\overline{D}^0\pi^+\pi^-$.  \label{fig:3}}
\end{figure}

\end{document}